\documentclass[article,a4,12pt]{article}
\usepackage{latexsym}
\usepackage{graphicx}

\title{Numerical simulation of the massive scalar field evolution in the Reissner-Nordstr\"{o}m black hole background}
\author{
Lihui Xue$^{1}$, Bin Wang$^{1}$ and Ru-Keng Su$^{2,1}$
}
\date{}

\begin{document}
\maketitle

\begin{center}
$^{1.}$ Department of Physics, Fudan University, Shanghai 200433,\\
People's Republic of China \\
$^{2.}$ China Center of Advanced Science and Technology (World Lab),\\
P. O. Box. 8730, Beijing 100080, People's Republic of China \\
\end{center}

\abstract{
We studied the massive scalar wave propagation in the background of Reissner-Nordstr\"{o}m black hole by using numerical simulations. We learned that the value $Mm$ plays an important role in determining the properties of the relaxation of the perturbation. For $Mm \ll 1$ the relaxation process depends only on the field parameter and does not depend on the spacetime parameters. For $Mm \gg 1$, the dependence of the relaxation on the black hole parameters appears. The bigger mass of the black hole, the faster the perturbation decays. The difference of the relaxation process caused by the black hole charge $Q$ has also been exhibited.\\
\\
PACS number(s): 04.20.Ex, 04.70.Bw
}

\section{Introduction}
The study of the evolution of various fields outside black holes plays an important role in black hole physics. 
In virtue of previous works, we now have the schematic picture regarding the dynamics of waves outside a spherical collapsing body. 
One intriguing subject is that a static observer outside the black hole can indicate a quasinormal ringing, 
which has frequencies directly relating to the parameters of the black hole. 
This quasinormal ringing is believed to be a unique fingerprint which would lead to the direct identification of the 
black hole existence and expected to be detected through gravitational wave observation in the near future \cite{01}. 
In order to extract as much information as possible from gravitational wave signal, 
it is of interest to understand quasinormal modes of various fields outside different black hole spacetimes.\\

The massless neutral external perturbations were first studied by Price \cite{02}. 
Studying the behavior of a massless scalar field propagating on a fixed Schwarzschild background, 
he showed that for an observer on a fixed position the field dies off with a power-law tail. 
The most complete picture to date on this problem was presented by Gundlach, Price, and Pullin \cite{03}. 
They found that the result given in \cite{02} also holds at the event horizon and the future null infinity. 
Similar results for a massless scalar field propagating on Reissner-Nordstr\"{o}m (RN) background have been 
obtained in \cite{04}\cite{05}\cite{06}. A close dependence of the quasinormal frequencies on the mass and 
charge of the RN black hole has been disclosed. The evolutions of charged massless scarlar field around a RN black hole and dilaton black hole have
been investigated in detail in \cite{Hod} and \cite{M}. 

A number of authors have studied radiative dynamics in black hole spacetimes that are not asymptotically flat. 
Brady, et al \cite{07} considered the evolution of a massless scalar field in Schwarzschild de Sitter and RN de Sitter spacetimes. 
They found that at late times the field decays exponentially, not as an inverse power-law. 
Motivated by the recent discovery of the AdS/CFT correspondence, the investigation of the quasinormal modes of AdS black holes becomes appealing. 
The analysis of the quasinormal modes for massless scalar field in four, five, and seven dimensional Schwarzschild Ads black holes was performed in \cite{08}. 
It was argued that the quasinormal frequencies have different properties concerning different size of black holes. 
This argument was confirmed by studying the small Schwarzschild AdS black holes \cite{09}. 
Extending the study of the evolution of the massless scalar field to RN AdS and topological black hole backgrounds \cite{10}\cite{11}, 
richer dependence of the quasinormal frequencies on the black hole parameters has been exhibited.\\

All these previous works are concentrated on massless scalar field perturbations. The evolution of massive scalar fields in black hole backgrounds is also of interest. Studying analytically the massive scalar field on a RN spacetime \cite{12} Hod and Piran argued that the intermediate asymptotic behavior of the field depends only on the field's parameter and does not have relation to the spacetime properties. Their result was obtained by expanding the wave equations for the massive scalar field in the black hole background as a power series in $M/r$, $Q/r$ and neglecting terms of order $O \left[ (Mm/r)^2 \right]$ and higher. Similar behavior was also claimed by Koyama et al in Schwarzschild black hole \cite{13} and Moderski et al in dilaton black hole cases \cite{14}. These results imply that one cannot get any information about the structure of the background spacetime from the intermediate asymptotic tails of massive scalar fields. It's quite different from the properties of the quasinormal modes of massless scalar field. Recent exact expressions of the quasinormal frequencies of massive scalar field for three-dimensional AdS \cite{15} and dS \cite{16} spacetimes got by either solving the wave equation in the bulk or studying the perturbation of conformal field theory on the boundary showed a contrary property to that of four-dimensional cases \cite{12}\cite{13}\cite{14}. The dependence of black hole parameters has been exhibited. Since the exact solution of quasinormal modes in four-dimensional spacetimes cannot be obtained analytically, it is of interest to carry out the numerical simulation of the collapse of a massive scalar field and compare with the asymptotic results got before. This is the motivation of the present paper.\\

The outline of this paper is as follows. In Sec. II we describe the physical system and formulate the evolution equations. We also discuss in detail the intermediate range and the approximations involved. In Sec. III we give the numerical results. Sec. IV will be a brief summary.\\

\section{Approximate Solution}
We consider the evolution of massive scalar fields in the Reissner-Nordstr\"{o}m black hole background described by the metric
\begin{equation}
ds^2=-\left(  1-\frac{2M}{r}+\frac{Q^2}{r^2} \right) dt^2+\left( 1-\frac{2M}{r}+\frac{Q^2}{r^2} \right)^{-1}dr^2+r^2d \Omega^2,
\label{metric}
\end{equation}
where $M$ and $Q$ correspond to the mass and charge of the black hole. The tortoise radial coordinate $y$ is defined by
\begin{equation}
dy=\frac{dr}{\lambda^2},
\end{equation}
where
\begin{equation}
\lambda^2=1-\frac{2M}{r}+\frac{Q^2}{r^2},
\end{equation}
Using the tortoise coordinate, Eq.(\ref{metric}) becomes
\begin{equation}
ds^2=\lambda^2(-dt^2+dy^2)+r^2d \Omega^2.
\end{equation}
The scalar field with mass $m$ satisfies the wave equation
\begin{equation}
(\Box-m^2)\phi=0.
\end{equation}
Resolving the field into spherical harmonics $\phi=\sum_{l,m}\psi_m^l(t,r)Y_l^m(\theta,\varphi)/r$, one obtains a wave equation for each multipole moment
\begin{equation}
\psi_{,tt}-\psi_{,yy}+V\psi=0,
\label{field_eq}
\end{equation}
where
\begin{equation}
V=\left( 1-\frac{2M}{r}+\frac{Q^2}{r^2} \right)\left[ \frac{l(l+1)}{r^2}+\frac{2M}{r^3}-\frac{2Q^2}{r^4}+m^2 \right].
\label{potential}
\end{equation}
The background spacetime parameters $M$ and $Q$ are obviously included in Eq.(\ref{potential}).\\
In order to study the time evolution of a massive scalar field, the retarded Green's function $G(y,y';t)$ is defined as \cite{12}
\begin{equation}
\left[ \frac{\partial^2}{\partial t^2}-\frac{\partial^2}{\partial y^2}+V(r) \right]G(y,y';t)=\delta(t)\delta(y-y'),
\end{equation}
and Eq.(\ref{field_eq}) can be written as
\begin{equation}
\psi(y,t)=\int \left[ G(y,y';t)\psi_t(y',0)+G_t(y,y';t)\psi(y',0) \right]dy'.
\end{equation}
The initial condition reads $G(y,y';t)=0$ for $t \le 0$. $G(y,y';t)$ can be found by using the Fourier transformation
\begin{equation}
\tilde{G}(y,y';\omega)=\int_{0^-}^{\infty}G(y,y';t)e^{i \omega t}dt.
\end{equation}
The Fourier component of the Green's function $\tilde{G}(y,y';\omega)$ can be expressed in terms of two linearly independent solutions $\tilde{\psi}_1(y,\omega)$ and $\tilde{\psi}_2(y,\omega)$ to the homogeneous equation
\begin{equation}
\left( \frac{d^2}{dy^2}+\omega^2-V \right)\tilde{\psi}_i(y,\omega)=0; i=1,2.
\end{equation}
Introduce an auxiliary variable $\xi$ in such a way that $\xi=\lambda \tilde{\psi}$ and remember $dy=dr/\lambda^2$ one can rewrite the equation of $\tilde{\psi}_i$ as
\begin{equation}
\frac{d^2\xi}{dr^2}+\left( \frac{\omega^2}{\lambda^4}-\frac{\frac{l(l+1)}{r^2}+\frac{2M}{r^3}-\frac{2Q^2}{r^4}+m^2}{\lambda^2}-\frac{1}{\lambda} \frac{d^2\lambda}{dr^2} \right)\xi=0.
\label{eq_of_xi}
\end{equation}
In order to calculate $\xi$, some approximations have been made in \cite{12} to simplify the equation. Assuming that the observer is situated far away from the black hole $(M \ll r)$, Eq.(\ref{eq_of_xi}) can be expanded in power series of $M/r$ and $Q/r$ to the order $O[M^2/r^2]$ and we arrive at
\begin{eqnarray}
&&\xi ''+\left[ \omega^2-m^2+\frac{-2Mm^2+4M\omega^2}{r}+\frac{-l(l+1)}{r^2} \right.\nonumber \\
&&\left.+\frac{M^2-Q^2}{r^4}+O[M^2/r^4] \right]\xi=0.
\label{xi_power}
\end{eqnarray}
Eq.(\ref{xi_power}) can be further approximated to the form
\begin{equation}
\xi ''+\left( \omega^2-m^2-\frac{l(l+1)}{r^2} \right)\xi=0,
\label{xi_inter}
\end{equation}
if $\frac{Mm^2}{r} \ll m^2$, $\frac{Mm^2}{r} \ll \frac{l(l+1)}{r^2}$, $\frac{M^2}{r^4} \ll m^2$, and $\frac{M^2}{r^4} \ll \frac{l(l+1)}{r^2}$. One can read from these four inequalities the intermediate range $\sqrt{\frac{M}{m}} \ll r \ll \frac{1}{Mm^2}$ in the limit $Mm \ll 1$. Eq.(\ref{xi_inter}) shows that the backscattering of the field from asymptotically far region has been neglected. From this equation, it was argued that the intermediate behavior of massive scalar field depends only on the field's parameter and it does not depend on the spacetime parameters \cite{12}.\\
With the increase of $Mm$, the intermediate range will shrink. When $Mm \ge 1$, the intermediate range will no longer exist.\\

\section{Numerical Simulation}
In this section we do not introduce any approximation and directly solve Eq.(\ref{field_eq}) numerically. It should be noted that Eq.(\ref{field_eq}) is invariant under the rescaling
\begin{equation}
r \rightarrow ar; t \rightarrow at; M \rightarrow aM; Q \rightarrow aQ; m \rightarrow m/a,
\end{equation}
where $a$ is a positive constant. In such a rescaling $Mm$ is invariant. Using the null coordinates $u=t-y$ and $v=t+y$, Eq.(\ref{field_eq}) can be recast as
\begin{equation}
4\psi_{,uv}+V(r)\psi=0.
\label{eq_of_uv}
\end{equation}
Eq.(\ref{eq_of_uv}) can be integrated numerically by the finite difference method suggested in \cite{03}. The late time evolution of a massive scalar field should not depend on the form of the initial data. In our numerical calculations, we use a Gaussian pulse of the form
\begin{equation}
\psi(u=0,v)=A \exp \{-\left[ \left( v-v_0 \right)/\sigma \right]^2 \}.
\end{equation}
After the integration is completed, the value $\psi(u_{max},v)$ is extracted, where $u_{max}$ is the maximum value of $u$ on the numerical grid. Taking sufficiently large $u_{max}$, $\psi(u_{max},v)$ represents a good approximation for the wave function at the event horizon. Since it has been shown that wave behavior is the same near or far away from the event horizon \cite{03}, we will study the dependence of wave behavior on background parameters in different ranges of $Mm$.\\
To test our numerical program, we first take $m=0$ in our simulation and investigate the dependence of the decay rate of the massless scalar field on the black hole charge. The results are shown in Fig. 1. There is a maximum of the slope at a critical value of the black hole charge $Q_{C}$. When $Q>Q_{C}$, the massless scalar field settles down slower as the increase of $Q$ until its extreme value $Q=M$. Considering the relation between the damping timescale $\tau$ and the imaginary part of the quasinormal frequencies $\omega_{I}$, $\tau=1/\left| \omega_{I} \right|$, our numerical result exhibits that $\left| \omega_{I} \right|$ has a maximum value for a critical value of the black hole charge. This wiggle behavior of $\left| \omega_{I} \right|$ is in accordance with the lowest normal mode frequencies studied in \cite{04}\cite{05}.\\

% begin figure 1 %
\begin{center}
\includegraphics[width=4in,height=3in]{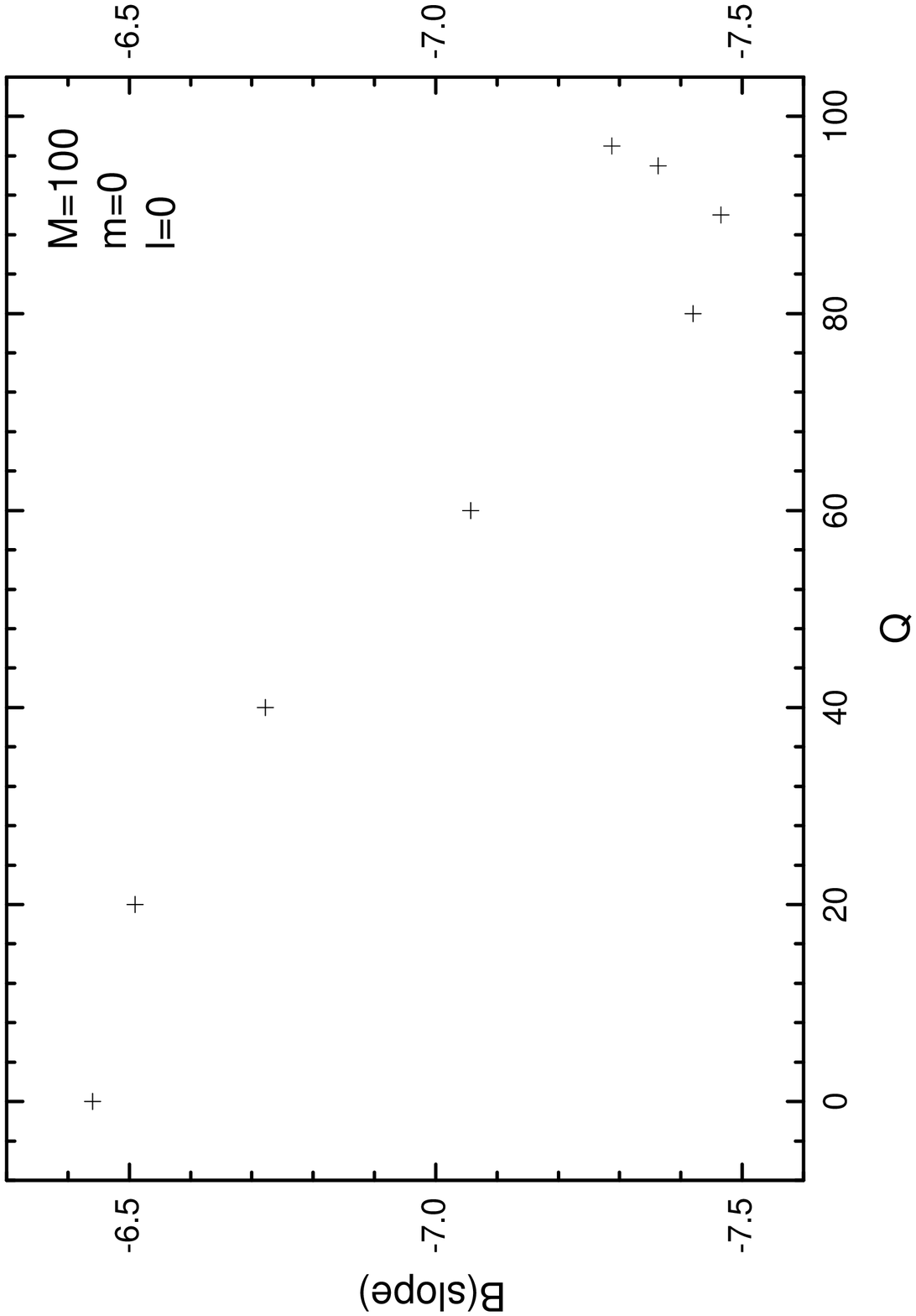}\\
\end{center}
\textit{Fig. 1. The decay rate for different black hole charges $Q=0$, $20$, $40$, $60$, $80$, $90$, $95$, $97$ when $l=0$. The black hole mass is taken to be $M=100$, and the mass of the scalar field is $m=0$.\\}
% end figure 1 % 

We now report our numerical simulation results of evolving massive scalar field on a RN black hole background. In the first series of numerical calculations, we choose appropriate values of the black hole mass $M$ and scalar field mass $m$ to satisfy $Mm \ll 1$. For different value of $M$, the results are shown in Fig. 2 with the multipoles $l=0$. In order to display the property of the relaxation of the perturbation clearly, we here just connected highest points of each wave crest and neglected the oscillations. It is obvious from this figure that in the region $Mm \ll 1$ the decay rate keeps approximately the same for different values of the black hole mass. The small difference in the slope to the order $10^{-2}$ was just caused by selecting points in getting the slope of the damping. This result also holds by taking different values of the charge of the black hole, which is different from that of the massless scalar field as shown in Fig. 1. The numerical pictures obtained here support the approximate solution in \cite{12}. The intermediate asymptotic behavior of massive scalar perturbation does not depend on the background black hole parameters.\\

% begin figure 2 %
\begin{center}
\includegraphics[width=4in,height=3in]{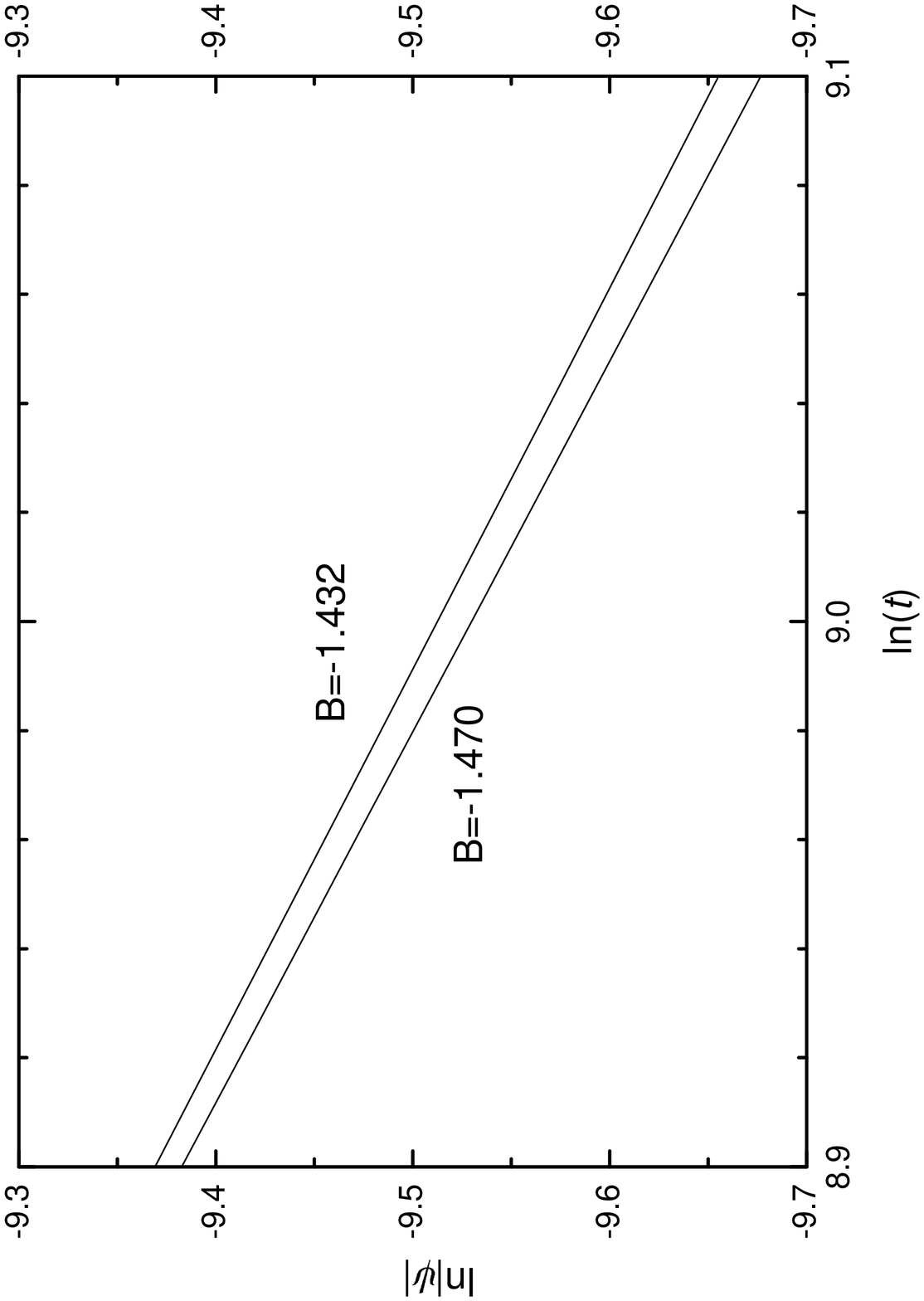}\\
\end{center}
\textit{Fig. 2a. Evolution of the massive field $|\psi|$ on the Reissner-Nordstr\"{o}m background, for $m=0.01$ and $l=0$. The field at a fixed radius ($r=50$) is shown as a function of time. The decay rate of the perturbation (the slope of the curve) $B$ is $-1.470$ for $M=0.5, Q=0.45$ (bottom curve) and $-1.432$ for $M=1.0, Q=0.45$ (top curve).\\}
\begin{center}
\includegraphics[width=4in,height=3in]{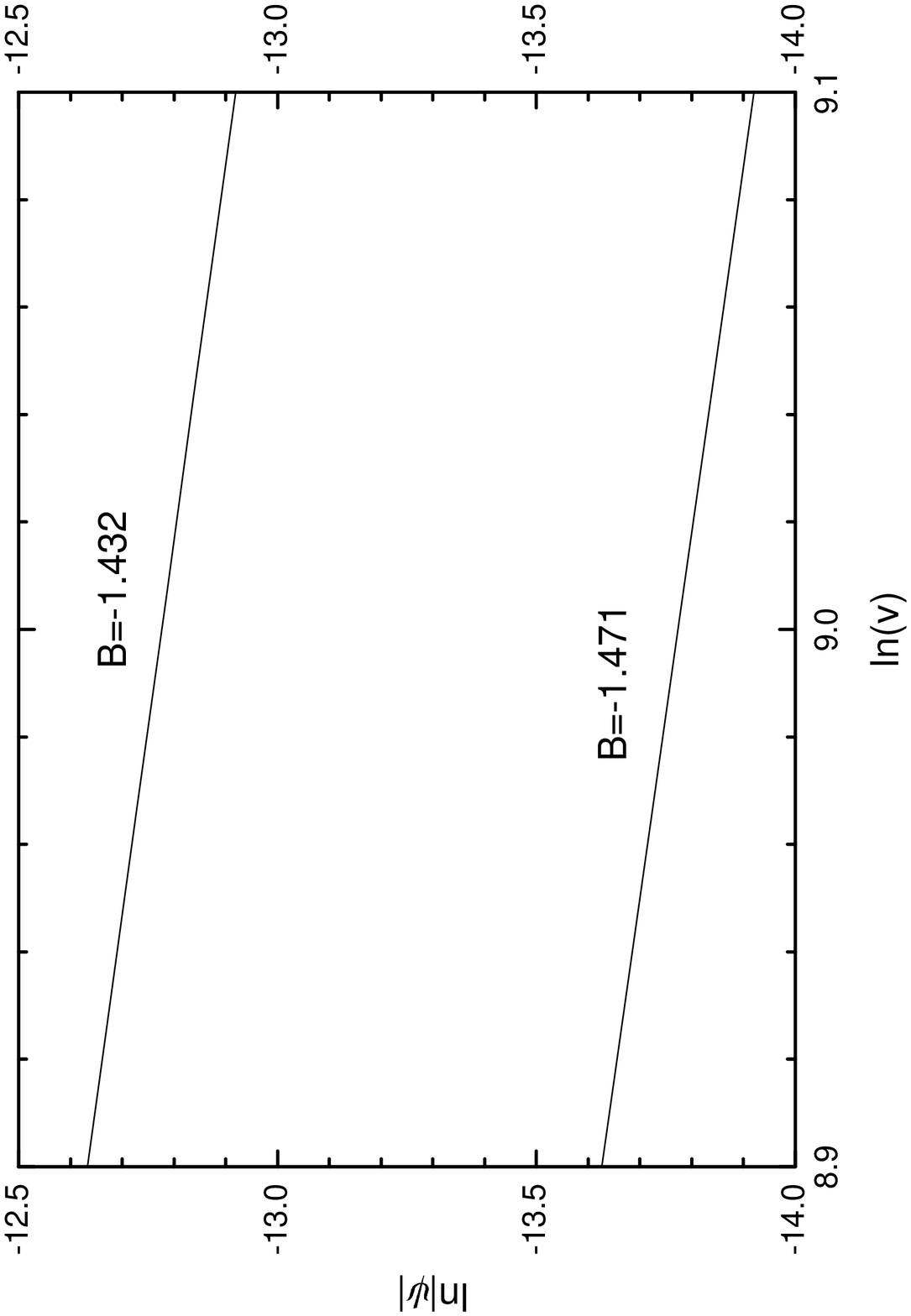}\\
\end{center}
\textit{Fig. 2b. The amplitude of the field along the event horizon for $m=0.01$ and $l=0$. The field is shown as a function of $v$. The black hole parameter is $M=0.5, Q=0.45$ for the bottom curve and $M=1.0, Q=0.45$ for the top curve. For $Mm \ll 1$, the decay rate for the field along the event horizon is almost identical to that at a fixed radius.\\}
% end figure 2 % 

The dependence of the decay rate of the perturbation on the black hole parameters appears for big value of $Mm$. For $Mm \gg 1$, we have found a difference of decay rate caused by the background parameters. Results displayed in Fig. 3 tell us that the bigger black hole mass leads to the faster decay of the massive scalar field perturbation. The difference of the relaxation process caused by the black hole charge $Q$ has also be exhibited in this region in Fig. 4 with the black hole mass fixed to be $M=200$. We see that the slope (the decay rate) of the scalar perturbation, first decreases with the increase of $Q$ and then over a critical value of $Q$, the slope increases with the increase of $Q$ until $Q$ reaches the extreme value. This behavior also holds for other values of black hole mass and for $M$ is bigger, the dependence is clearer. This means that before some critical value of $Q$, the larger black hole charge is, the slower the outside perturbation dies out, corresponding to the decrease of the imaginary quasinormal frequency; while over the critical value of $Q$, we have the opposite decay behavior with the increase of the imaginary frequency. This wiggle behavior of the imaginary quasinormal frequency caused by black hole charge arise in the massive scalar field perturbation is qualitatively different from the case of the massless scalar field.

% begin figure 3 %
\begin{center}
\includegraphics[width=4in,height=3in]{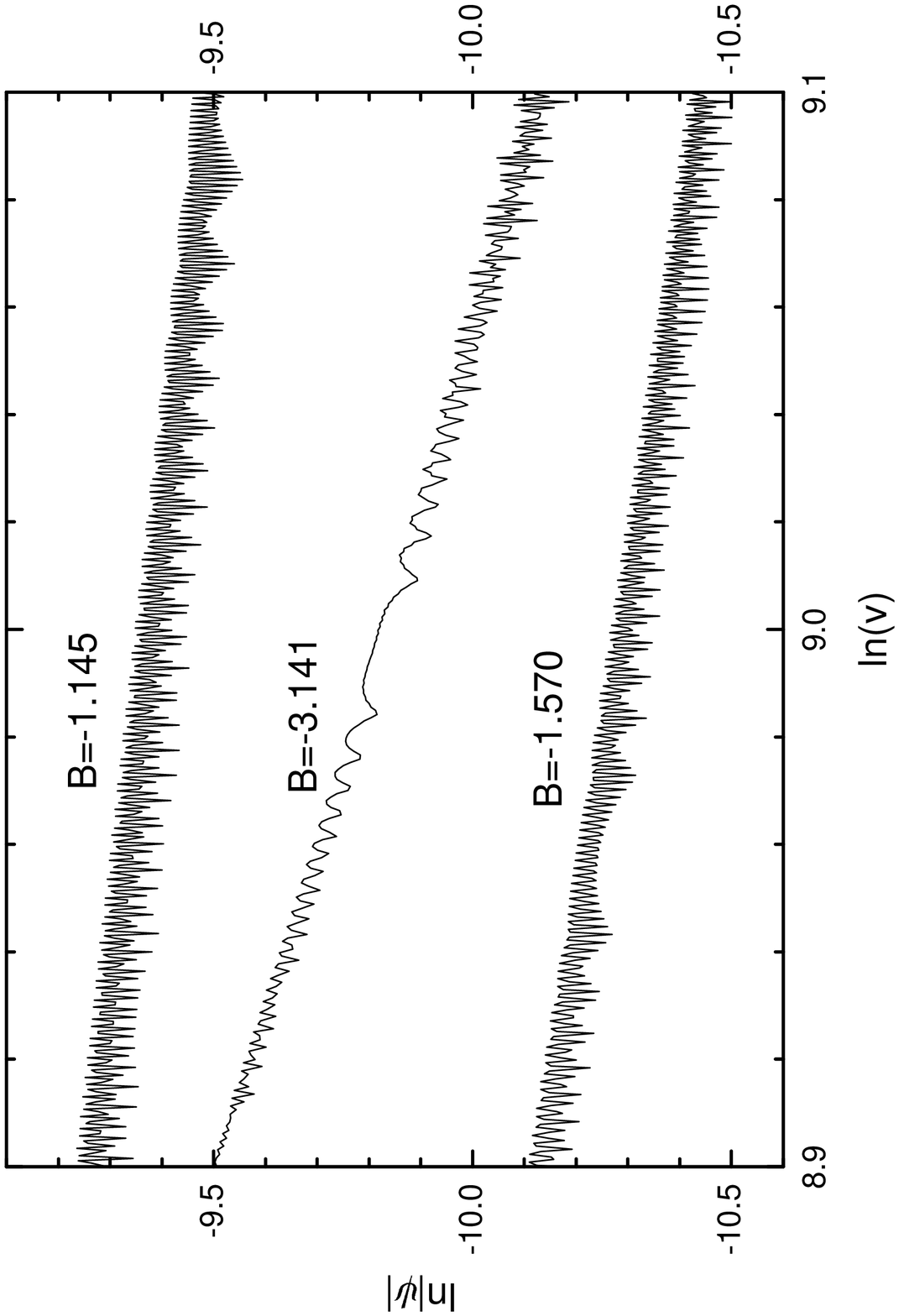}\\
\end{center}
\textit{Fig. 3. Evolution of the massive field on the RN background for $Mm \gg 1$. The field along the event horizon is shown as a function of $v$. From top to bottom, the black hole mass and charge are $M=50, Q=45$; $M=200, Q=45$, and $M=100, Q=45$, respectively. The field mass is $m=1.0$ and the multipole index is $l=0$.\\}
% end figure 3 %

% begin figure 4 %
\begin{center}
\includegraphics[width=4in,height=3in]{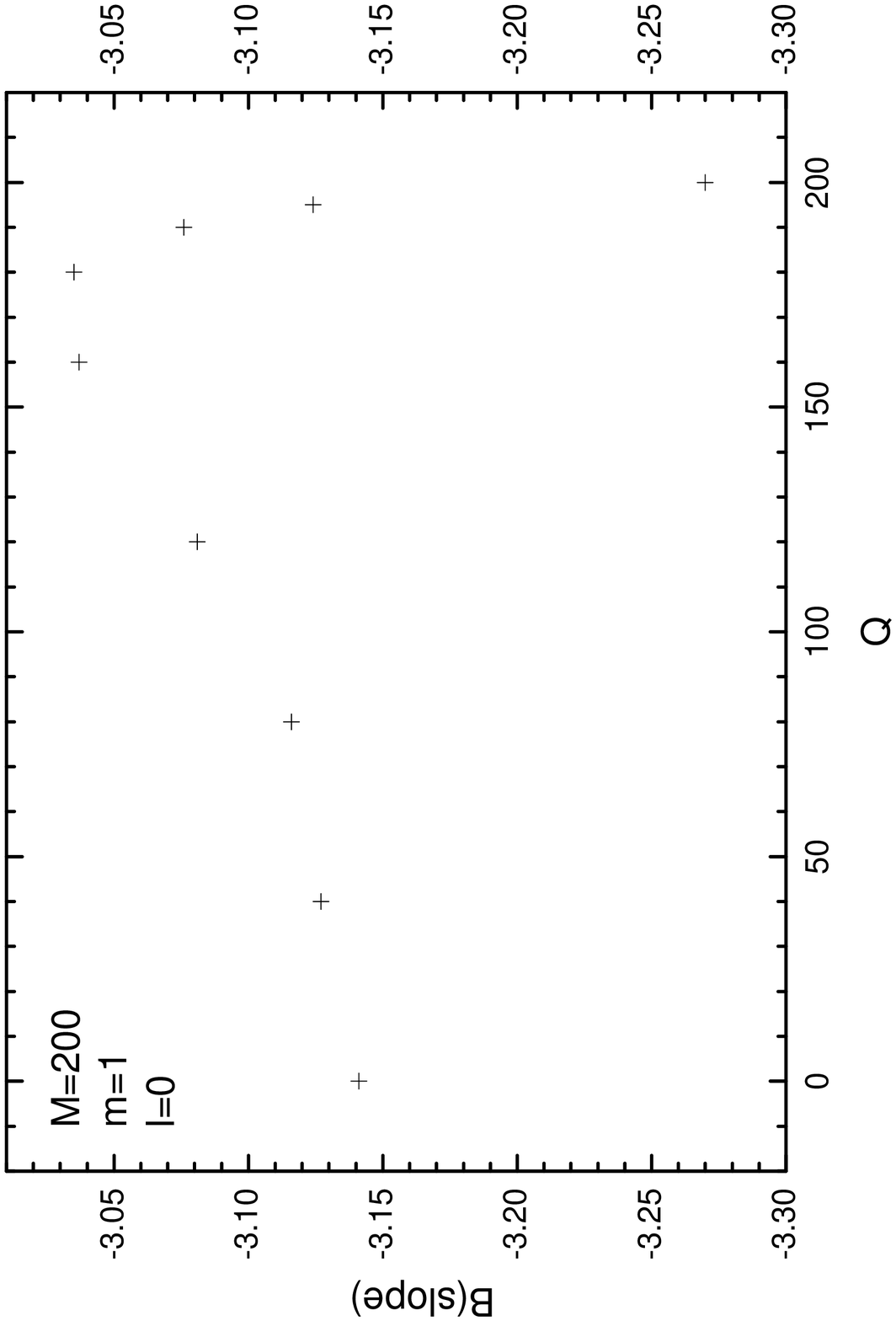}\\
\end{center}
\textit{Fig. 4. The decay rate for different black hole charge $Q=0$, $40$, $80$, $120$, $160$, $180$, $190$, $195$ and $200$. The black hole mass is taken to be $M=200$, and  $m=1.0$, $l=0$.\\}
% end figure 4 %

\section{Conclusions}
We have studied the massive scalar wave propagation in the background of RN black hole by using numerical simulations instead of solving the wave equation approximately. The field evolution behavior we found is different from that of massless scalar field. We have learnt that the value $Mm$ plays an important role in determining the properties of the relaxation of the perturbation. For $Mm \ll 1$, our numerical result confirmed the approximate argument given in \cite{12}, that the intermediate behavior of massive scalar perturbations on a black hole background approximately depends only on the field's parameter and it does not depend on the spacetime parameters. For $Mm \ge 1$, the intermediate approximation condition breaks down. Our numerical results show that for $Mm>1$, the dependence of the relaxation of the perturbation on the background black hole mass appears. This kind of dependence becomes much clearer for $Mm \gg 1$. In the region $Mm \gg 1$, the influence given by the black hole charge $Q$ on the massive scalar field's relaxation has also been observed. For small values of $Q$, the settle down of massive scalar field perturbation is slower for larger charge. However, after the black hole charge reaches a critical value, we got an opposite behavior: the field decays faster with the increase of the black hole charge. The picture of the massive scalar field relaxation caused by the black hole charge is different from that of the massless scalar field. New phenomenon found here are quite interesting and the inclusion of the scalar field mass have enriched the spectrum of the wave dynamics outside the black hole.\\
\\
Acknowledgements: This work was partially supported by NNSF of China. We would like to acknowledge the helpful discussions with Weigang Qiu.\\

\end{document}